\documentclass[12pt]{article}
\usepackage{makeidx,showidx}
\usepackage{latexsym}
\usepackage{amsmath}
\usepackage{amssymb}
\usepackage{varioref}
\usepackage{epsfig}
\usepackage{epsf}
\usepackage{subfigure}
\usepackage{rotating}
\usepackage{amscd}
\usepackage{cite}
\makeatletter
\newsavebox{\tempbox}
\renewcommand{\@makecaption}[2] {
  \vspace{10pt}\sbox{\tempbox}{{\footnotesize \sc #1 --} #2} 
  \ifthenelse{\lengthtest{\wd\tempbox} > \linewidth}
    { {\footnotesize \sc #1 --} #2\par }
      {\begin{center}{\footnotesize \sc #1 --} #2 \end{center}} }
\makeatother
\newcommand{\close}{\emph{close}}
\newcommand{\far}{\emph{far}}
\newcommand{\zompa}{\vspace{ .5 cm}}

\newcommand{\nutau}{\nu_\tau}
\newcommand{\numu}{\nu_{\mu}}
\newcommand{\nux}{\nu_x}
\newcommand{\dmq}{{\Delta m}^2}
\newcommand{\evq}{{\mathrm{eV}}^2}
\newcommand{\sint}{\sin^2 2\theta}
\newcommand{\CERN}{{CERN}}
\newcommand{\FNAL}{{FNAL}}
\newcommand{\skam}{Super--Kamiokande}
\begin{document}
\title{About testing $\numu$\ oscillation with $\dmq$\ smaller than 0.001 $\evq$\ with the \CERN\ Proton Synchrotron}
\author{P. F. Loverre, R. Santacesaria, F. R. Spada \\ {\small \it Universit\`a ``La Sapienza'' and Istituto Nazionale di Fisica Nucleare (INFN)} \\{\small \it  Rome, Italy} \\ -- \\ {\small Submitted to \emph{The European Physical Journal C}}}
\date{}
\maketitle
\begin{abstract}
We study the feasibility of a long--baseline neutrino experiment from \CERN\ to Gran Sasso LNGS Laboratories using the \CERN\ PS accelerator. Baseline and neutrino energy spectrum are suitable to explore a region of the ($\dmq$, $\sint$) parameters space which is not reached by K2K, the first experiment that will test at accelerator the atmospheric neutrino anomaly put in evidence by \skam.
\end{abstract}

The recent \skam\ measurements of atmospheric neutrino fluxes~\cite{skam:1998} favour $\numu \rightarrow \nutau$ (or $\numu \rightarrow \nux$) oscillations, with almost maximal mixing and $\dmq$\ in the range $(5 \div 60) \cdot 10^{-4} \, \evq$.

The first test of this atmospheric neutrino anomaly at accelerator will be performed in Japan by the K2K~\cite{k2k:1997} experiment. 
K2K has recently star\-ted taking data using a neutrino beam generated by the KEK 12--GeV Proton Synchrotron directed toward the \skam\ detector, which is placed about 250 Km away from KEK.
The K2K experiment, owing to an L/E ratio of order 250/1 (Km/GeV), explores via disappearance $\numu$\ oscillations down to $\dmq \sim 2 \cdot 10^{-3} \, \evq$. 

The same $\dmq$\ region can be explored with higher sensitivity with the high energy neutrino beams of \FNAL\ (NuMI~--~MINOS experiment~\cite{minos:web, ables:1995}) and \CERN\ (\CERN~--~Gran Sasso LNGS beam~\cite{cern-gs:1998}).
The higher energy will also provide the opportunity for a direct search of $\nutau$ charged current interactions. 

\zompa

However, if the K2K experiment will not find any attenuation or distortion of the $\numu$\ flux, it will become of primary importance to perform an accelerator experiment capable to extend the oscillation search to lower values of $\dmq$, down to the limit $\dmq = 5 \cdot 10^{-4} \, \evq$\ of the region suggested by the atmospheric measurements of \skam.

The use of a low energy neutrino beam from the Cern Protosynchrotron (PS) allows this measurement. 

The importance of an experiment based on a neutrino beam from the PS has already been underlined by F. Dydak et al.~\cite{dydak:nufact}. In that paper, however, the accent has been put on a long term, detailed exploration of the oscillation mechanism; the project would be based, already in a first phase, on a major upgrade of the PS machine, allowing to improve the intensity by a factor 100.

In our opinion, the existing machine could directly be exploited to test $\dmq$\ values down to $5 \cdot 10^{-4} \, \evq$. The intensity of the PS is in fact high enough to make possible a long--baseline, \CERN~--~Gran Sasso, $\numu$\ disappearance experiment. A neutrino beam from the PS would have an energy similar to that of the K2K beam, while the baseline of the PS experiment would be 732 Km instead of 250 Km.
If it is possible to collect at Gran Sasso a statistics similar to that of the K2K experiment, the factor 3 in distance converts directly in a sensitivity to 3 times smaller $\dmq$, i.e. $\dmq \simeq 6~\cdot~10^{-4} \, \evq$.

In a relatively short term, the only other project 
which could allow to study oscillations with $\dmq$ below 0.001 $\evq$ 
is the use by MINOS of a low energy beam, the PH2(low) option described in ref.~\citen{minos:web}.
We have not made a detailed comparison of that
beam with the PS beam. We just note that the characteristics of 
the two beams are quite different. The 
Fermilab low energy beam has higher intensity, 
but the energy spectrum is peaking at somewhat higher energies
(2--3 GeV compared to the 1.2 GeV of the PS beam)
with significant contributions up to 50 GeV. These
features might complicate the analysis of the data
with respect to the case of the PS beam.
\begin{figure}[tb] \begin{center} \mbox{ \epsfig{file=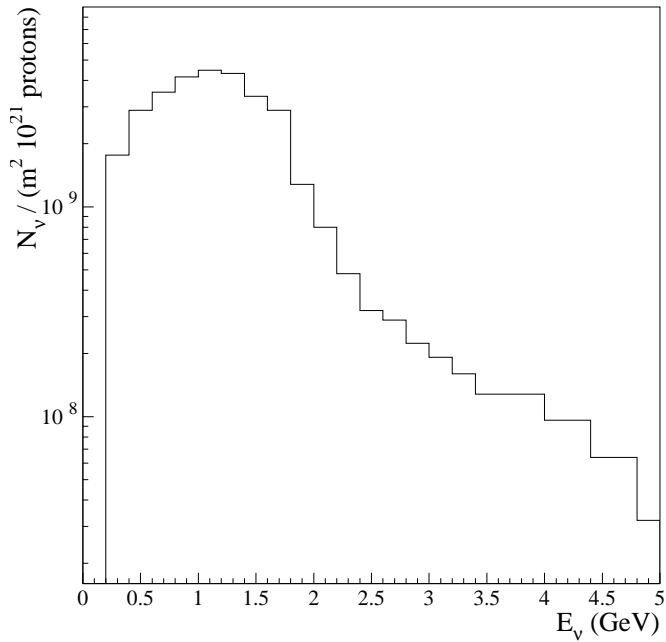,width=.7\textwidth} } \caption{\footnotesize \it The $\numu$\ flux from the \CERN\ Protosynchrotron (PS), assuming a decay tunnel length of 300~m, a baseline of 732~Km and $10^{21}$ protons on target.} \label{fig:flux} \end{center} \end{figure}

Our calculations of neutrino flux, event rates and corresponding sensitivity to neutrino oscillations for a PS \CERN~--~Gran Sasso experiment are presented in the following.

\zompa

We have simulated the PS neutrino beam in the setup used by the BEBC~--~PS~180~\cite{bebc:1986} experiment (proton energy of 19.2 GeV and horn focusing). Assuming a decay tunnel length of 300 m, we obtain at a distance of 732 Km the $\numu$\ flux displayed in fig.~\ref{fig:flux} for $10^{21}$ protons on target.

An experiment designed to search for $\numu$\ disappearance would have to measure the absolute rate and the energy spectrum of $\numu$\ charged current interactions, and compare them with the prediction based on the measurement performed in a similar detector located close to the neutrino source.
In our evaluation of the sensitivity to oscillation of the experiment, the event rates have been calculated assuming, as in ref.~\citen{i216:1997}, a cross section of $0.85 \cdot 10^{-38} \, \mathrm{cm}^2$ for the quasi--elastic reaction $\numu \, n \rightarrow \mu^- \, p$.
In the simulation, $\numu \, n \rightarrow \mu^- \, p$ reactions represent $50 \%$ of the total number of charged current interactions. 

The effect of a finite experimental resolution was taken into account by imposing the energy cut $E_\nu > 0.6$ GeV. We assume that with this cut it will be possible to identify and measure events with a $\mu$ in the final state with 100\% efficiency for quasi--elastic interactions and with 50\% efficiency for the remaining charged current interactions.
Event rates were then computed for
\renewcommand{\labelitemi}{\bfseries --}
\begin{itemize}
\item a total flux of $1.0 \cdot 10^{21}$ protons on target (POTs)
\item a fiducial mass of 20 Ktons for the detector.
\end{itemize}
\noindent We will justify these two assumptions later.
\begin{figure}[tb] \begin{center} \mbox{ \epsfig{file=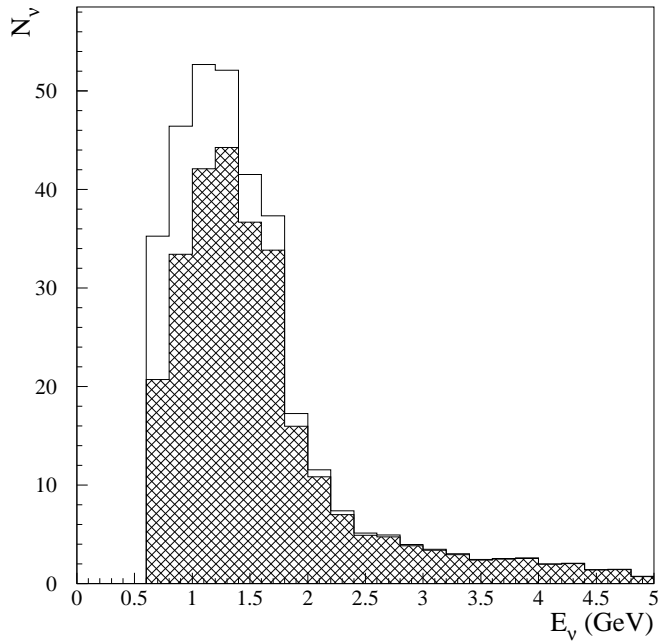,width=.7\textwidth} } \caption{\footnotesize \it Energy spectrum of the interactions with a muon in the final state for a target mass of 20 KTons, assuming a 100\% efficiency on quasi--elastic interactions and a 50\% efficiency on the remaining charged current interactions, and the cut $E_\nu > 0.6$ GeV. The hatched inset shows the spectrum resulting from a $\numu \rightarrow \nutau$ oscillation with $\sint=1$ and $\dmq=7 \cdot 10^{-4} \, \evq$.} \label{fig:mu} \end{center} \end{figure}

The resulting energy spectrum of the interactions with a muon in the final state is shown in fig.~\ref{fig:mu}. 
\begin{figure}[tb] \begin{center} \mbox{ \epsfig{file=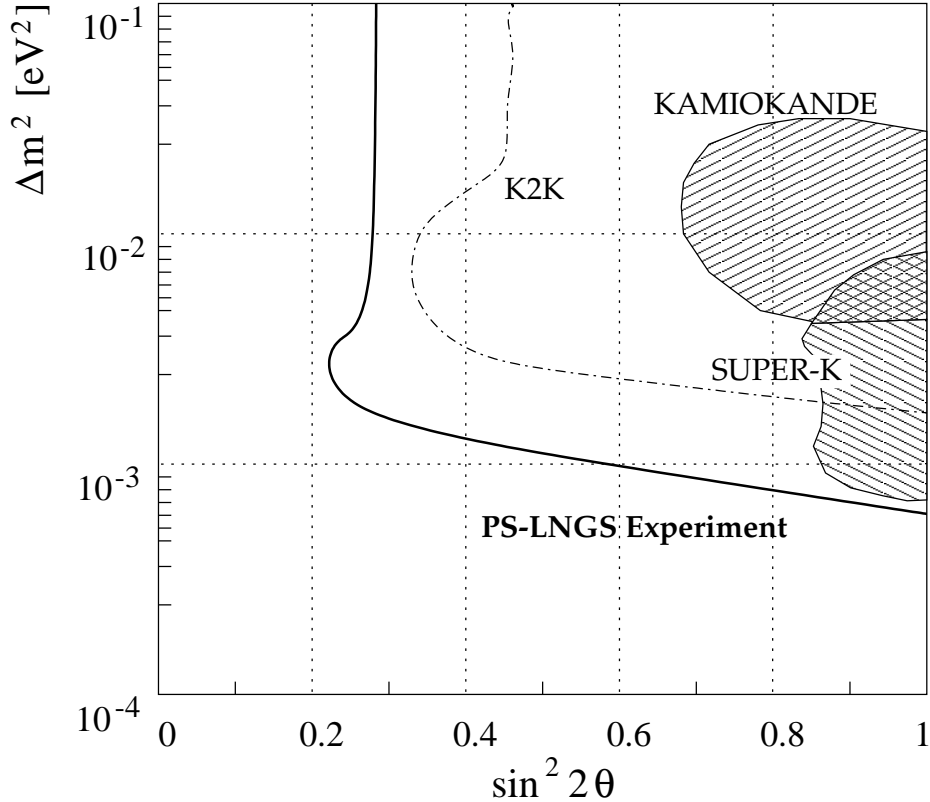,width=.9\textwidth} } \caption{\footnotesize \it Sensitivity to $\numu$\ oscillations for a long baseline PS -- LNGS experiment. Also shown are the design sensitivity of the K2K experiment and the result obtained by the measurements on atmospheric neutrinos by the Kamiokande~\cite{kamio:1998ea} and \skam~\cite{skam:1998} experiments.} \label{fig:osc} \end{center} \end{figure}
The histogram contains 337 events. If 
oscillations occurred, their 
effect would be a reduction of the number of interactions and,
for $\dmq < 0.01 \, \evq$, a distortion of the energy 
spectrum. We assume that the total rate at the \far\ detector
can be predicted with a 10\% systematic error from a
measurement at the \close\ detector. Fig.~\ref{fig:osc} then shows the
90\% $C.L.$ exclusion curve which could be obtained in absence
of oscillation, adding quadratically systematic and
statistical errors. The whole region of the parameter space
suggested by the Kamiokande and \skam\ measurements would be covered.

Of course, the experiment should aim at detecting oscillation with more than 1.28 $\sigma$ significance. This will be possible
because at low $\dmq$\  the oscillation would also
produce a distortion of the energy 
spectrum. As an example, the inset 
in fig.~\ref{fig:mu} shows the energy spectrum
of the interacting $\numu$\ neutrinos for the case of 
a $\numu \rightarrow \nutau$ oscillation with 
$\sint=1$ and $\dmq=7 \cdot 10^{-4} \, \evq$.
With respect to the no-oscillation spectrum, there
are 76 events less, and 50 of them are missing from the
first three bins of the histogram, yielding an 
evidence for oscillation of more than three standard deviations.
\zompa

We conclude that, in spite of the limited statistics, the experiment would be capable to detect oscillations in the whole range of the ($\dmq$, $\sint$) parameters suggested by \skam.

The experiment will need a very large \far\ detector, a similar, but much smaller \close\ detector, and a very intense use of the PS.

Our calculations assumed an integrated flux corresponding to $1.0 \cdot 10^{21}$ POTs. Presently, the radiation constraints limit the PS intensity to an average value of $1.2 \cdot 10^{13}$ protons on target per second; therefore, four years of data taking with a detector of 20 Ktons would be required. This is a very heavy but achievable task; a better investigation of the beam design -- optimisation of beam energy, target and horn design -- may well lead to a substantial improvement of the neutrino flux and to a consequent reduction of running time and/or detector mass.

If no evidence for oscillation will be found by the K2K experiment, the discussed option deserves great attention.

\end{document}